\documentclass[12pt]{article}

\usepackage{amsfonts,amssymb,amsmath}
\usepackage{graphicx}
\DeclareGraphicsExtensions{epsfig}
\begin{document}
\title{\bf Quantum speed limit time for a two-qubit system interacting with independent and common reservoirs  }
\author{ N. Behzadi $^{a}$
\thanks{E-mail:n.behzadi@tabrizu.ac.ir}  ,
A. Ektesabi $^{b}$ ,
B. Ahansaz $^{b}$,
E. Faizi $^{b}$
\\ $^a${\small Research Institute for Fundamental Sciences, University of Tabriz, Iran,}
\\ $^b${\small Physics Department, Azarbaijan shahid madani university, Iran.}} \maketitle

\begin{abstract}
\noindent
We study the quantum speed limit time for a two-qubit system interacts with independent and common reservoir. The system is initially prepared in a class of X-structure density matrix, namely the extended Werner-like states (EWL). We demonstrate that when reservoir enters to the non-Markovian regime, the quantum speed limit time decrease more sharply in a common reservoir than independent reservoirs.
\\
\\
{\bf PACS Nos:}
\end{abstract}

\section{Introduction}
Quantum mechanics imposes fundamental limit to the speed of quantum evolution by time-energy uncertainty relation: $\tau\geq max\{\pi\hbar/(2\Delta E),\pi\hbar/(2E)\}$, where $\Delta E$ is the variance of the energy and $E$ its average  energy with respect to ground state. This bounds known as Mandelstam-Tamm $\cite{Mandelstam}$ and Margolus-Levitin $\cite{Margolus}$ bounds. By combining this bounds, we can determine the maximum rate of evolution for closed quantum systems $\tau_{QSL}= max\{\pi\hbar/(2\Delta E),\pi\hbar/(2E)\}$, the so-called quantum speed limit time
which is the minimum time required for a quantum system to transform to a different state.

Since any system is inevitable subjected to an environment, quantum speed limit time bound for open quantum systems is highly desirable. In single qubit open systems, quantum speed limit time bound have been described in terms of the operator norm  of the nonunitary generator of the dynamics with using a geometric approach which provided by the Bures angle $\cite{Jozsa,Bures}$,  between initial and final states of the quantum system $\cite{Deffner}$. They extended both Mandelstam-Tamm and Margolus-Levitin type bounds to open quantum systems and showed that the non-Markovian effects can speed up quantum evolution and subsequently leads to smaller quantum speed limit time. However their quantum speed limit time bounds is derived from pure initial states and is not attainable for mixed states. Another quantum speed limit time bound in open quantum sytem is derived by using of quantum Fisher information for time estimation. However for deriving this bound it is necessary  to minimize the Fisher information in the enlarged system-environment space, which is too hard to evaluate in the case of initially mixed states $\cite{taddei}$. Recently Zhe Sun $et$ $al$. $\cite{Zhe Sun}$ derived a new quantum speed limit bound for open quantum systems by using an alternative fidelity as a distance measure, which is computable and the initial states can be chosen as either pure or mixed states.  For a two-qubit system that each qubit interacts with independent reservoirs and initially is prepared in pure states (Bell-type states), quantum speed limit time have been investigated in Ref $\cite{Chen}$, and it was also found that the non-Markovian effect of environment, can speedup the quantum evolution and shortening the quantum speed limit time.

 In this paper, we study the quantum speed limit time for a two-qubit system interacts with independent and common reservoirs while initially is prepared in a class of X-structure density matrix, namely the extended Werner-like states (EWL). We show that the quantum speed limit time has a sharper bound when system interacts with a common reservoir and in transition from Markovian regime to non-Markovian regime decreases steeply in a common reservoir than independent reservoirs. So we can see the evolution of a two-qubit system is faster and the non-Markovian effect of reservoir is significant when two-qubit system interacts with a common reservoir.

 In the following sections, we first introduce the quantum speed limit time bound, then we will investigate the quantum speed limit time for independent reservoirs to study  the Markovian and non-Markovian effects of reservoir and in the following section we will do this investigation for common reservoirs and compare the results of quantum speed limit time in common and independent reservoirs. Finally, the paper will end by a brief conclusion.

\section{Quantum speed limit time}
In this section, we use the quantum speed limit time bound derived in $\cite{Zhe Sun}$ to evaluate the speed of evolution of a typical two-qubit system interacting with independent and common reservoirs. To this aim, we exploit an alternative fidelity definition, as the distance measure of two quantum states, introduced in $\cite{Wang}$ as
\begin{eqnarray}
F(\rho_{0},\rho_{t})=\frac{Tr(\rho_{0}\rho_{t})}{\sqrt{Tr(\rho_{0}^{2})Tr(\rho_{t}^{2})}},
\end{eqnarray}
to evaluate the QSL time bound, as obtained in $\cite{Zhe Sun}$ (note that $\rho(t)\equiv\rho_{t}$). According to the Ref. $\cite{Zhe Sun}$, the derived QSL time bound which is applicable to either Markovian and
non-Markovian dynamics, is as the following form
\begin{eqnarray}
\tau\geq\tau_{QSL}=\frac{|1-F(\rho_{0},\rho_{\tau})|}{X(\tau)},
\end{eqnarray}
where
\begin{eqnarray}
X(\tau)=\frac{2}{\tau}\int_{0}^{\tau}\sqrt{\frac{Tr(\dot{\rho}_{t}^{2})}{Tr(\rho_{t}^{2})}}dt,
\end{eqnarray}
by denoting that $\dot{\rho}_{t}$ is the time derivative of the state $\rho_{t}$ and $\tau$ is the actual driving time. The initial state can be chosen either pure or mixed, which is one of the superiority of this bound to the others from the computability point of view. In the next section, we investigate the dynamics of the two-qubit system interacting with independent and common reservoirs and compare the obtained results on the speed of evolution of the system.

\section{Independent reservoirs}
We consider a bipartite two-qubit system each of them interacts with a zero-temperature bosonic reservoir. The Hamiltonian for each qubit that coupled to their own reservoir can be written as
\begin{eqnarray}
H=\frac{1}{2}\hbar\omega_{0}\sigma_{z}+\sum_{k}\hbar\omega_{k}a_{k}^\dag a_{k}+\sum_{k}\hbar(g_{k}a_{k}\sigma_{+}+g_{k}^*a_{k}^\dag \sigma_{-}),
\end{eqnarray}
where, $\sigma_{z}$ is the Pauli matrix and $\sigma_{+}$ ($\sigma_{-}$) is the Pauli raising (lowering) operator for the atom with transition frequency $\omega_{0}$. $a_{k}$ ($a_{k}^\dag$) is the annihilation (creation) operator for the $k$th field mode with frequency $\omega_{k}$ and $g_{k}$ is the coupling constant between the $k$th field mode and the atom. We consider the quantum evolution given by a dynamical completely positive trace preserving (CPT) map $\Phi(t,0)$ which maps the initial state $\rho(0)$ to the state $\rho(t)$ at time $t\geq0$,
\begin{eqnarray}
\rho(0)\rightarrow\rho(t)=\Phi(t,0)\rho(0).
\end{eqnarray}
The dynamics of a single qubit system has an exact solution $\cite{Breuer}$, and leads to a dynamical map $\Phi(t,0)$
which can be shown as a reduced density matrix $\rho(t)$ as follows
\begin{eqnarray}
\rho(t)=\begin{pmatrix}
|G(t)|^2\rho_{11}(0)&G(t)\rho_{10}(0)\\
G^{*}(t)\rho_{01}(0)&\rho_{00}(0)+(1-|G(t)|^2)\rho_{11}(0)\\
\end{pmatrix},
\end{eqnarray}
where the $ \rho_{ij} (t)=\langle i|\rho_{s}(t)|j\rangle $ denotes the matrix elements of $\rho_{s}(t)$. The function $G(t)$ is define as the solution of the equation
\begin{eqnarray}
\frac{d}{dt}G(t)=-\int_{0}^{t} dt_{1}f(t-t_{1})G(t_{1}),
\end{eqnarray}
corresponding to the initial condition $G(0)=1$, where the kernel $f(t-t_{1})$ is called the correlation function,
\begin{eqnarray}
\begin{array}{c}
f(t-t_{1})=\langle0|B(t)B^{\dag}(t)|0\rangle e^{i\omega_{0}(t-t_{1})}\\\\\sum_{k}|g_{k}|^2 e^{i(\omega_{0}-\omega_{k})(t-t_{1})},
\end{array}
\end{eqnarray}
where
\begin{eqnarray}
B(t)=\sum_{k}g_{k}b_{k}e^{-i\omega_{k}t},
\end{eqnarray}
is the environmental operator. This kernel may be expressed in terms of the spectral density $J(\omega)$ of the reservoir as follows
\begin{eqnarray}
f(t-t_{1})=\int d \omega J(\omega)e^{i(\omega_{0}-\omega_{k})(t-t_{1})}.
\end{eqnarray}
In this paper we use the exactly solvable damped Jaynes-Cumming model for a two-level resonantly coupled to a dissipative single mode reservoir $\cite{Breuer}$ and with considering the reservoir as a Lorantzian spectral density
\begin{eqnarray}
J(\omega)=\frac{1}{2\pi}\frac{\gamma_{0}\lambda}{(\omega-\omega_{0})^2+\lambda^2},
\end{eqnarray}
where $\gamma_{0}$ is the coupling strength and $\lambda$ is the width of the Lorentzian function which is related to environmental correlation time by $\tau_{E}=\lambda^{-1}$,
we can find the correlation function as
\begin{eqnarray}
 f(\tau)=\frac{1}{2}\gamma_{0}e^{-\lambda|\tau|},
\end{eqnarray}
 Solving Eq. (7) with the correlation function (12) we find
\begin{eqnarray}
G(t)=e^{-\lambda t/2} [\cosh(\frac{dt}{2})+\frac{\lambda}{d} \sinh(\frac{dt}{2})],
\end{eqnarray}
where $d=\sqrt{\lambda^{2}-2\gamma_{0}\lambda}$.

Now we can determine the dynamics of a two-qubit system which each of them interacts with an independent reservoir by using the dynamical map given by $\Phi^{AB}(t)=\Phi^{A}(t)\otimes\Phi^{B}(t)$. By applying this map to the initial states, we can obtain dynamics for a bipartite two qubit system in the independent reservoir. Here, we want to investigate quantum speed limit time for both initially pure and mixed entangled states. For this purpose, we considering a class of X-structure density matrix, namely the extended Werner-like states (EWL)$\cite{Munro,Wei}$, defined as
\begin{eqnarray}
\rho_{EWL}^{\Psi_{1}}=r|\Psi_{1}\rangle\langle\Psi_{1}|+\frac{1-r}{4}I,
\end{eqnarray}
where $|\Psi_{1}\rangle=\alpha|01\rangle+e^{1\theta}(1-\alpha^2)^{1/2}|10\rangle$, and
\begin{eqnarray}
\rho_{EWL}^{\Psi_{2}}=r|\Psi_{2}\rangle\langle\Psi_{2}|+\frac{1-r}{4}I,
\end{eqnarray}
where $|\Psi_{2}\rangle=\alpha|00\rangle+e^{1\theta}(1-\alpha^2)^{1/2}|11\rangle$,
$r$ is the purity parameter, $I$ is a $4\times4$ identity matrix.
For $r=0$, extended Werner-like states (EWL) become totally mixed states, while for $r=1$ they reduce to the Bell-like states of $|\Psi_{1}\rangle$ and $|\Psi_{2}\rangle$.
For initial state (14) with $\theta=0$, the reduced density matrix of system has the form:
\begin{eqnarray}
\begin{array}{c}
\rho_{11}(t)=\frac{1-r}{4}|G(t)|^4,\\\\
\rho_{22}(t)=\frac{r-1}{4}|G(t)|^4+\frac{(1-2\alpha^2)r+1}{2}|G(t)|^2,\\\\
\rho_{23}(t)=r\alpha\sqrt{1-\alpha^2}|G(t)|^2,\\\\
\rho_{32}(t)=r\alpha\sqrt{1-\alpha^2}|G(t)|^2,\\\\
\rho_{33}(t)=\frac{r-1}{4}|G(t)|^4+\frac{(2\alpha^2-1)r+1}{2}|G(t)|^2,\\\\
\rho_{44}(t)=\frac{1-r}{4}|G(t)|^4-|G(t)|^2+1.
\end{array}
\end{eqnarray}.

And for initial state (15) with $\theta=0$, the reduced density matrix of system is given by:
\begin{eqnarray}
\begin{array}{c}
\rho_{11}(t)=\frac{1+(3-4\alpha^2)r}{4}|G(t)|^4,\\\\
\rho_{14}(t)=r\alpha\sqrt{1-\alpha^2}G^2(t),\\\\
\rho_{22}= \rho_{33}= \frac{(4\alpha^2-3)r-1}{4}|G(t)|^4+\frac{(1-2\alpha^2)r+1}{2}|G(t)|^2,\\\\
\rho_{41}= r\alpha\sqrt{1-\alpha^2}G^{*^2}(t),\\\\
\rho_{44}=\frac{1+(3-4\alpha^2)r}{4}|G(t)|^4+((2\alpha^2-1)r-1)|G(t)|^2+1.
\end{array}
\end{eqnarray}
Now with using Eq. (2), we can plot quantum speed limit time bound for a two-qubit system each part interacting with individual independent reservoirs while initially prepared in extended Werner-like states. In Figs. 1(a) and 2(a) we show the quantum speed limit time bound for different initial state of Eqs. (14) and (15) with different mixed coefficient $r$, where two qubits interact with two independent reservoir. According to the Ref. $\cite{Breuer}$, $\gamma_{0}<\lambda/2$ corresponds to the weak coupling regime, where the behaviour of the system-environment is Markovian and $\gamma_{0}>\lambda/2$ where the non-Markovian dynamics occurs. As is shown in this figures, the quantum speed limit time bounds decreases when the reservoir enters to the non-Markovian regime. However the decreasing slope of the quantum speed limit time bounds are not steep and the non-Markovian effect of reservoir is not significant.

\section{Common reservoirs}
In this section we study the quantum speed limit time for a two-qubit system which contains at most two excitation, interacting with a common reservoir.
By considering a two-qubit system interacting with a common zero-temperature bosonic reservoir, the Hamiltonian of system in the basis ${|00\rangle,|01\rangle,|10\rangle,|11\rangle}$, is written by $H=H_{0}+H_{int}$, which reads
\begin{eqnarray}
H_{0}=\frac{1}{2}\hbar\omega_{0}(\sigma_{z}^A+\sigma_{z}^B)+\sum_{k}\hbar\omega_{k}a_{k}^\dag a_{k},
\end{eqnarray}
\begin{eqnarray}
H_{int}=(\sigma_{+}^A+\sigma_{-}^B)\sum_{k}g_{k}a_{k}+H.c.,
\end{eqnarray}
where $A$ and $B$ denote the two qubit interacting with a common reservoir. Like the independent reservoir situation we considering that two atoms interact resonantly with a Lorentzian reservoir, such as electromagnetic field with its initial state is the vacuum state. The Hamiltonian in the basis of${|0\rangle=|00\rangle,|+\rangle=(|10\rangle+|01\rangle)/2,|-\rangle=(|10\rangle+|01\rangle)/2,|2\rangle=|11\rangle}$ is given by
\begin{eqnarray}
H_{0}=2\omega_{0}|2\rangle\langle2|+\omega_{0}(|+\rangle\langle+|+|-\rangle\langle-|)+\sum_{k}\omega_{k}a_{k}^\dag a_{k},
\end{eqnarray}
\begin{eqnarray}
H_{int}=\sum_{k}\sqrt{2}g_{k}a_{k}(|+\rangle\langle0|+|2\rangle\langle+|)+H.c.,
\end{eqnarray}
where $|+\rangle$ and $|-\rangle$ are respectively, the super-radiant and subradiant states. We can describe this bipartite system by a four-state system which three states are coupled to the vacuum in a ladder configuration, and one state is decoupled from the other states and from the field, as we see from Eqs. (20) and (21), the super-radiant state is coupled to states $|0\rangle$ and $|2\rangle$ through the reservoir, while subradiant state does not decay. So we can consider the Hamiltonian of system as two parts, one describes the free dynamics of the subradiant $|-\rangle$ state and other part describes a three-state ladder system containing ${|+\rangle,|0\rangle,|1\rangle}$ states. In Ref $\cite{ Mazzola,Mazzolab}$ the dynamics of the three-level ladder system is exactly solved by  using of the pseudomode approach $\cite{ Garraway, Dalton}$. The pseudomode approach is based on the idea of extending the system to include part of reservoir (the pseudomode) to construct the bigger system for applying the Markov approximation. This method allow us to derive a Markovian master equation for the extended system. This master equation describes interaction of atom with pseudomodes which leaks in to a Markovian reservoir $\cite{Mazzola}$. In order to solve the dynamics of system, we assume that the two qubits interact with only one pseudomode which is associated with the Lorentzian spectral distribution as
\begin{eqnarray}
J(\omega)=\frac{\gamma_{0}^2}{2\pi}\frac{\Gamma}{(\omega-\omega_{0})^2+(\Gamma/2)^2},
\end{eqnarray}
where $\gamma_{0}$ is the coupling strength and $\Gamma$ is the width of the Lorentzian function. Therefore we can derive an exact Markovian master equation of extended system  as
\begin{eqnarray}
\frac{\partial\tilde{\rho}}{\partial t}=-i[V,\tilde{\rho}]-\frac{\Gamma}{2}(a^{\dag}a\tilde{\rho}+\tilde{\rho}a^{\dag}a-2a\tilde{\rho}a^{\dag}),
\end{eqnarray}
with
\begin{eqnarray}
V=\sqrt{2}\gamma_{0}(a|+\rangle\langle0|+a^{\dag}|0\rangle\langle+|+a|2\rangle\langle+|+a^{\dag}|+\rangle\langle2|),
\end{eqnarray}
where $\tilde{\rho}$ is the density matrix of the extended system, $a$ ($a^{\dag}$) is the annihilation (creation) operator. The master equation in Eqs.(23) and (24) are solved in Ref.$\cite{Mazzolab}$ and
the analytic solution of reduces density matrix for initial states of Eqs. (14) and (15) is presented in Appendix of $\cite{ Mazzolab}$. Now with using Eq. (2), we can derive the quantum speed limit time for a two-qubit system interacting with a common reservoir while is initially prepared in (EWL) states. The Markovian and non-Markovian regime can be distinguish by the relation of $ \Gamma$ and $\gamma_{0}$. The Markovian regime corresponds to the case $\gamma_{0}<\Gamma/4$ and the non-Markovian regime occurs in $\gamma_{0}>\Gamma/4$ $\cite{Mazzola}$. In Figs. 1(b) and 2(b) we show the quantum speed limit time bounds for different initial states of (14) and (15), where two-qubit system interacts with a common reservoir. As is clearly shown, the quantum speed limit time bounds decreases more steeply when reservoir enters to the non-Markovian regime. A comparison between the effects of common and independent reservoir on the quantum speed limit time bounds reveals that the quantum speed limit bounds are more sharper and the non-Markovian effect of reservoir is significant in speeding up of quantum evolution, when two-qubit interacts with a common reservoir.

\section{Conclusions}
We have investigated the quantum speed limit time for a two-qubit system interacting with a common and also with two independent non-Markovian reservoirs. For a class of (EWL) initially states, we have demonstrated that when reservoir enters to the non-Markovian region, the quantum speed limit time bound decrease more steeply when a two-qubit system interacts with a common reservoir. Thus we conclude that the evolution of system is faster and the non-Markovian effect of reservoir is significant when interacts with a common reservoir.

\newpage

Fig. 1: Quantum speed limit time $ \tau_{QSL}$ as a function of the coupling strength $\gamma_{0}$ (in units of $\omega_{0}$) for a two-qubit system with the initial state of (14) when (a) interacts with two independent reservoir and (b) a common reservoir. Parameters are $\lambda=50$ (in units of $\omega_{0}$), $\alpha=\frac{1}{\sqrt{2}}$, $\theta=0$ and  $\tau=1$.

\begin{figure}
\qquad \qquad\qquad\qquad \qquad a \qquad\qquad \quad\qquad\qquad\qquad\qquad\qquad\qquad b\\{
        \includegraphics[width=3in]{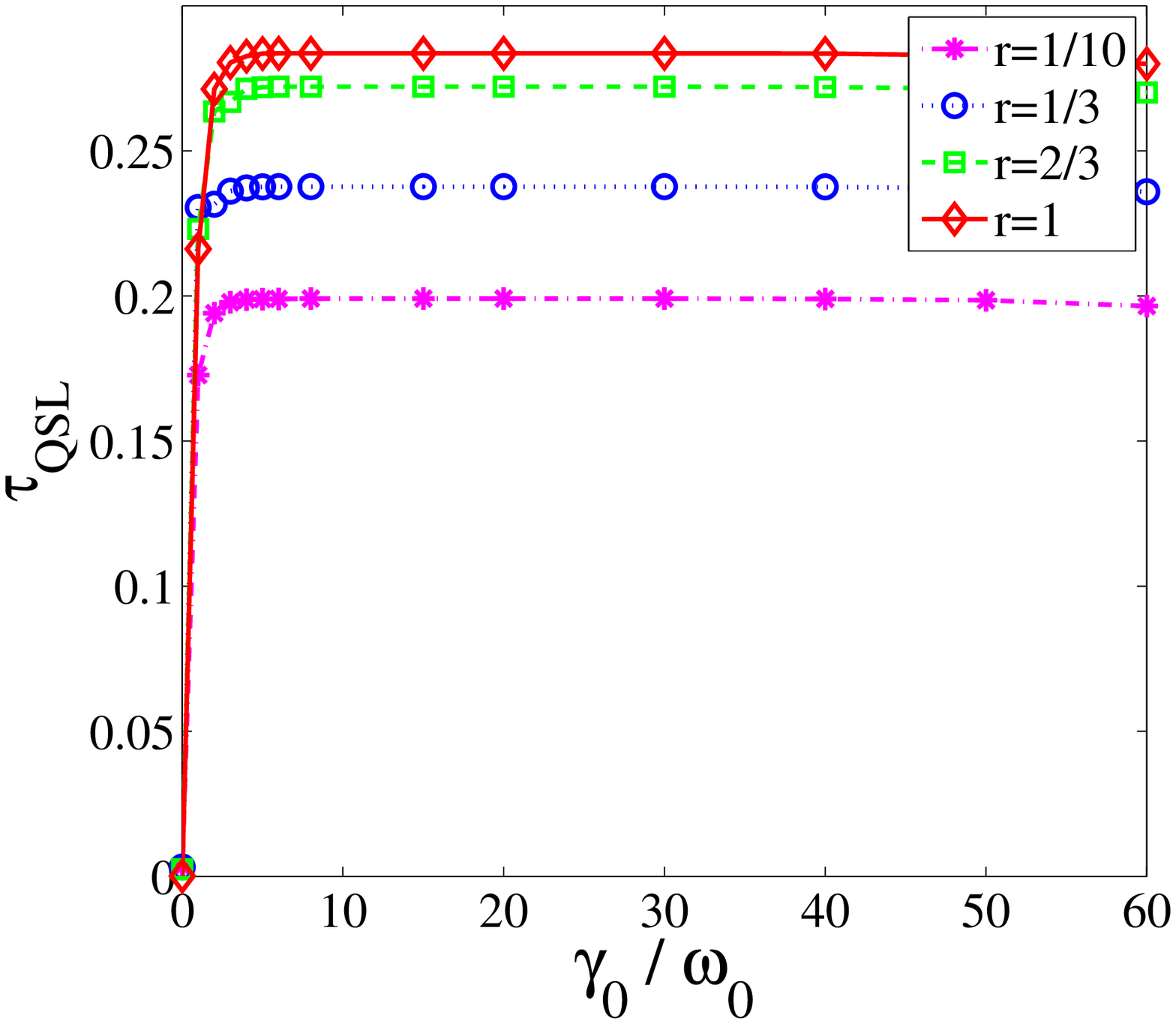}
        \label{fig:first_sub}
    }{
        \includegraphics[width=3in]{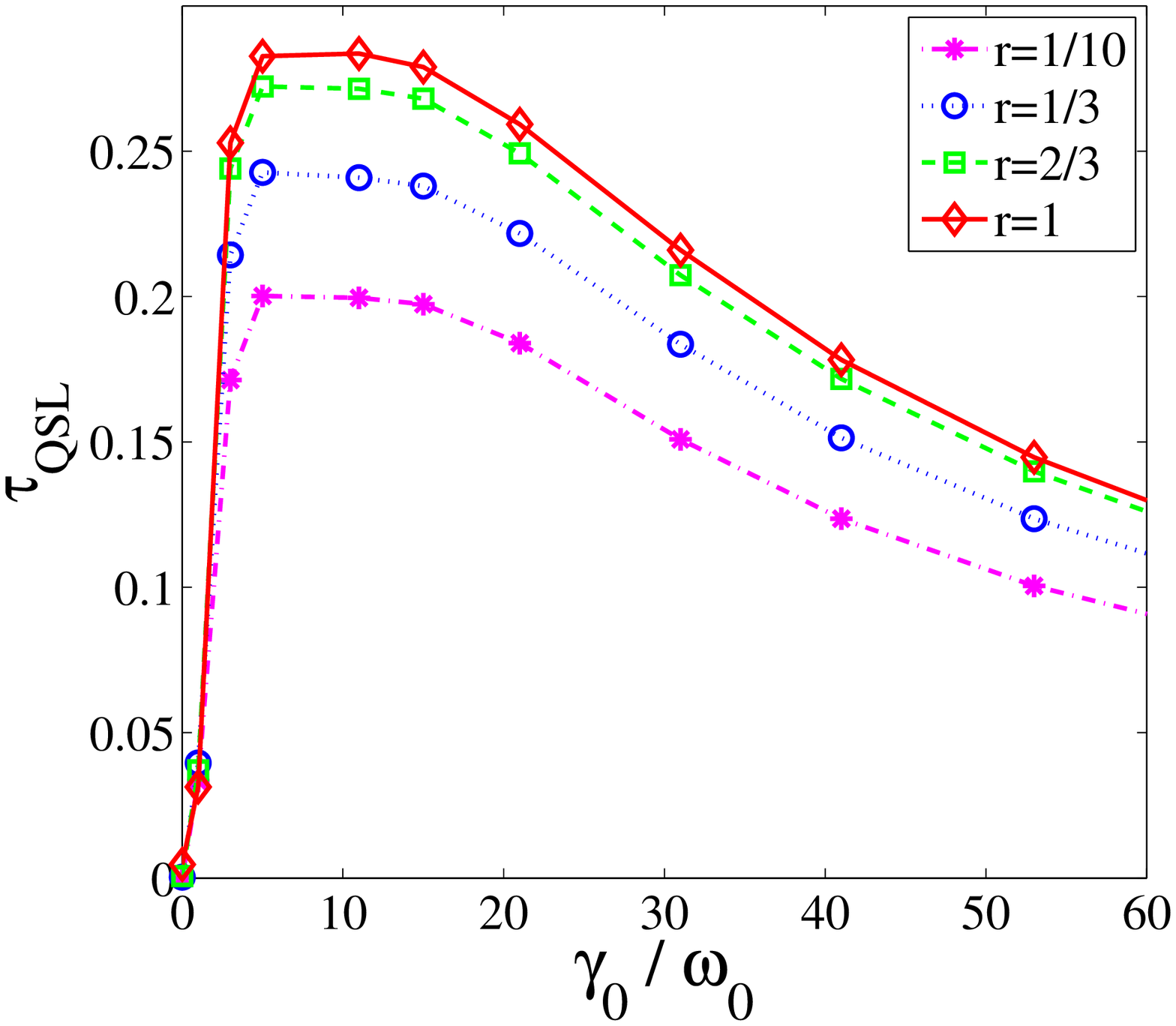}
        \label{fig:second_sub}
    }
    \caption{}
\end{figure}

\newpage
Fig. 2: Quantum speed limit time $ \tau_{QSL}$ as a function of the coupling strength $\gamma_{0}$ (in units of $\omega_{0}$) for a two-qubit system with the initial state of (15) when (a) interacts with two independent reservoir and (b) a common reservoir. Parameters are $\lambda=50$ (in units of $\omega_{0}$), $\alpha=\frac{1}{\sqrt{2}}$, $\theta=0$ and  $\tau=1$.

\begin{figure}
\qquad \qquad\qquad\qquad \qquad a \qquad\qquad \quad\qquad\qquad\qquad\qquad\qquad\qquad b\\{
        \includegraphics[width=3in]{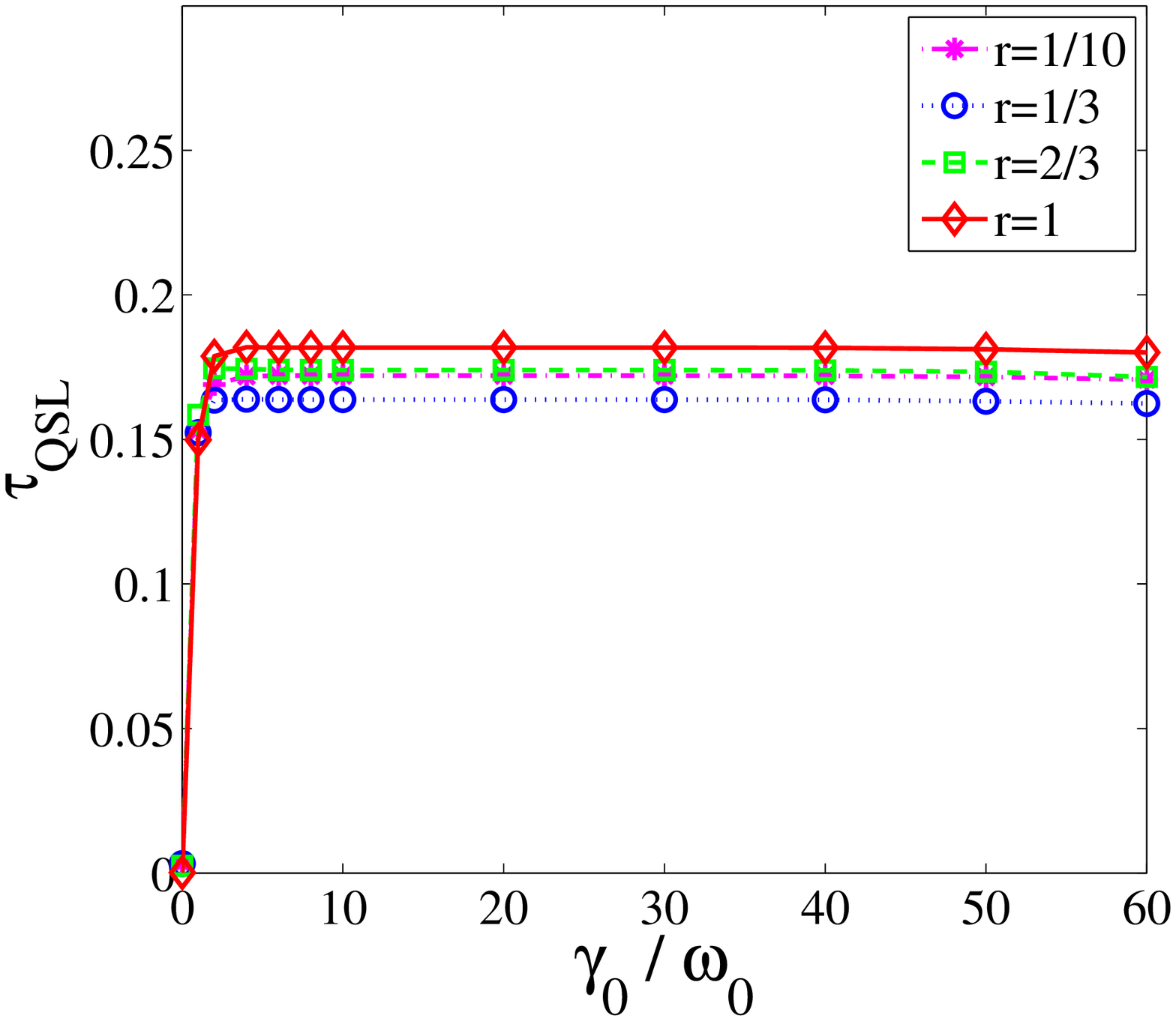}
        \label{fig:first_sub}
    }{
        \includegraphics[width=3in]{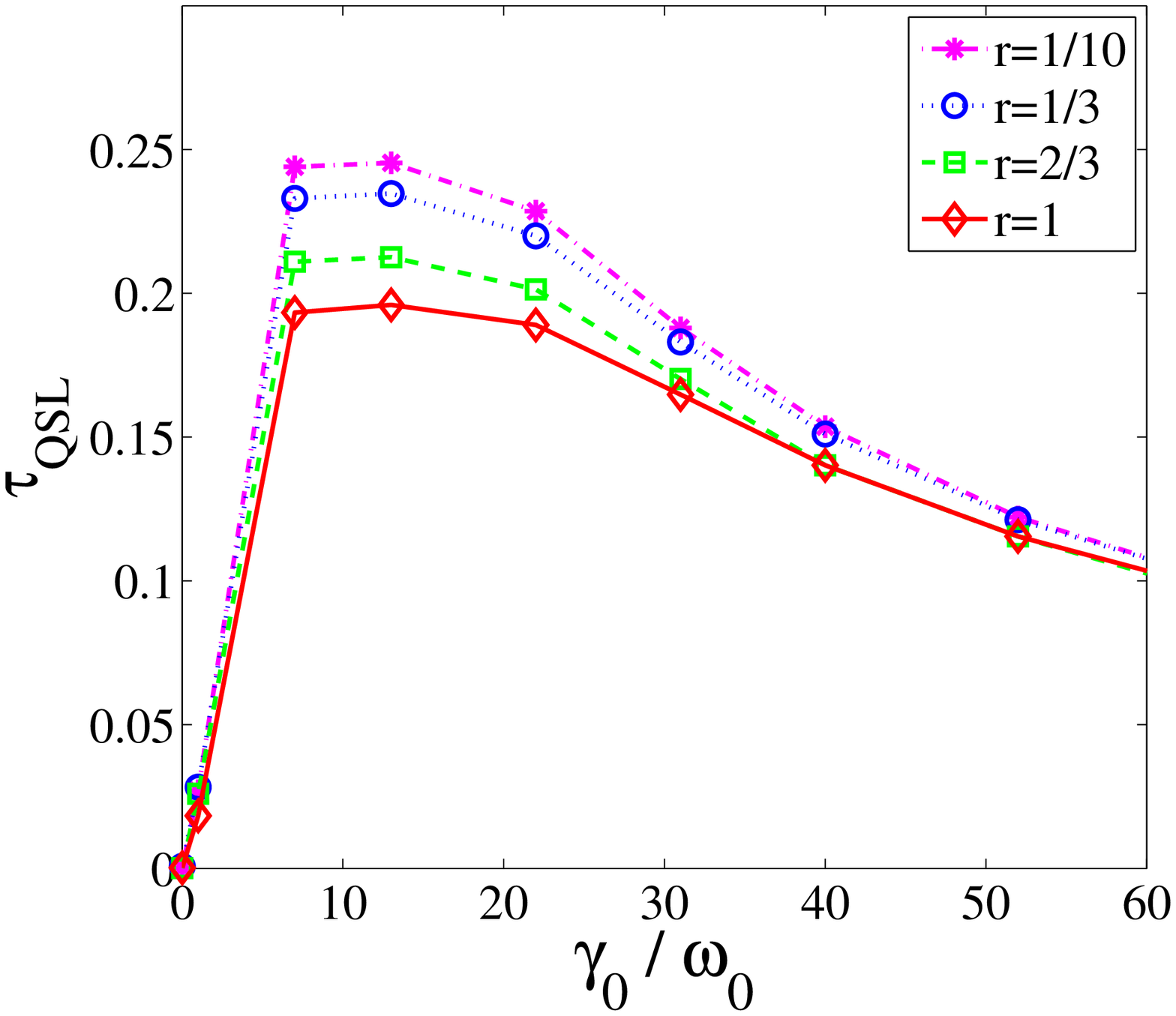}
        \label{fig:second_sub}
    }
    \caption{}
\end{figure}

\end{document}